\documentclass[prb,twocolumn,amsmath,amssymb,floatfix,superscriptaddress,aps]{revtex4-1}
\usepackage[utf8]{inputenc}
\usepackage{graphicx}
\usepackage{dcolumn}
\usepackage{bm}
\usepackage{xcolor}
\usepackage{soul}
\usepackage{cancel}

\usepackage{lipsum}
\usepackage{comment}
\usepackage{braket}
\usepackage{mathalfa}
\newcommand{\be}{\begin{equation}}
\newcommand{\ee}{\end{equation}}
\newcommand{\bea}{\begin{eqnarray}}
\newcommand{\eea}{\end{eqnarray}}
\newcommand{\ba}{\begin{align}}
\newcommand{\ea}{\end{align}}

\begin{document}

\title{Thermoelectric transport within density functional theory}

\author{Nahual Sobrino}
 \email{nahualcsc@dipc.org}
\affiliation{Donostia International Physics Center, Paseo Manuel de Lardizabal 4, E-20018 San Sebasti\'an, Spain}
\affiliation{Nano-Bio Spectroscopy Group and European Theoretical Spectroscopy Facility (ETSF), Departamento de Pol\'imeros y Materiales Avanzados: F\'isica, Qu\'imica y Tecnolog\'ia, Universidad del Pa\'is Vasco UPV/EHU, Avenida de Tolosa 72, E-20018 San Sebasti\'an, Spain}

\author{Florian Eich}
\affiliation{HQS  Quantum  Simulations  GmbH,  Haid-und-Neu-Straße  7,  D-76131  Karlsruhe,  Germany}

\author{Gianluca Stefanucci}
\affiliation{Dipartimento di Fisica, Universit\`{a} di Roma Tor Vergata,
Via della Ricerca Scientifica 1, 00133 Rome, Italy; European Theoretical
Spectroscopy Facility (ETSF)}
\affiliation{INFN, Laboratori Nazionali di Frascati, Via E. Fermi 40,
00044 Frascati, Italy}

 \author{Roberto D'Agosta}
\email{roberto.dagosta@ehu.es}
\affiliation{Nano-Bio Spectroscopy Group and European Theoretical Spectroscopy Facility (ETSF), Departamento de Pol\'imeros y Materiales Avanzados: F\'isica, Qu\'imica y Tecnolog\'ia, Universidad del Pa\'is Vasco UPV/EHU, Avenida de Tolosa 72, E-20018 San Sebasti\'an, Spain}
\affiliation{IKERBASQUE, Basque Foundation for Science, Plaza de Euskadi 5, E-48009 Bilbao, Spain}

\author{Stefan Kurth}
\email{stefan.kurth@ehu.es}
\affiliation{Nano-Bio Spectroscopy Group and European Theoretical Spectroscopy Facility (ETSF), Departamento de Pol\'imeros y Materiales Avanzados: F\'isica, Qu\'imica y Tecnolog\'ia, Universidad del Pa\'is Vasco UPV/EHU, Avenida de Tolosa 72, E-20018 San Sebasti\'an, Spain}
\affiliation{IKERBASQUE, Basque Foundation for Science, Plaza de Euskadi 5, E-48009 Bilbao, Spain}
\affiliation{Donostia International Physics Center, Paseo Manuel de Lardizabal 4, E-20018 San Sebasti\'an, Spain}

\date{\today}
\begin{abstract}
A new formalism to describe steady-state electronic and thermal transport in the framework of density functional
theory is presented. A one-to-one correspondence is proven between the three basic variables of the theory, i.e.,
the density on as well as the electrical and heat currents through the junction, and the three basic potentials,
i.e., the local potential in as well as the DC bias and thermal gradient across the junction. Consequently, the
Kohn-Sham system of the theory requires three exchange-correlations potentials. In linear response, the new
formalism leads to exact expressions for the many-body transport coefficients (both electrical and thermal
conductances and Seebeck coefficient) in terms of both the corresponding Kohn-Sham coefficients and derivatives
of the exchange-correlations potentials. The theory is applied to the Single Impurity Anderson Model, and an
accurate analytic parametrization for these derivatives in the Coulomb blockade regime is constructed through
reverse engineering.

\end{abstract}

\maketitle
\section{Introduction}
 
Thermoelectricity \cite{Goldsmid:10} is at the heart of a range of technological applications, e.g., energy
conversion, and is intrinsically related to both charge and heat transport. With progress in the
manipulation and fabrication of new materials at the nanoscale and even at the level of single molecules
(``Molecular Electronics'', see
Refs.~\onlinecite{me-book1,me-book2,ThossEvers:19,EversKorytarTewariRuitenbeek:20,dutta2020single}), designing more efficient
thermoelectrics requires reliable modelling techniques at an atomistic level. Today, density functional theory
(DFT) is most often the method of choice for ab-initio modelling due to its favorable balance of accuracy and
numerical efficiency. In a DFT framework, transport is typically described by combining DFT with the
Landauer-B\"uttiker (LB) approach. This LB-DFT formalism, also known as DFT-NEGF (DFT plus Nonequilibrium Green
Functions), treats (ballistic) transport as a scattering problem of non-interacting electrons. The resulting
Landauer formula for the electronic current is physically very intuitive in that the current is given as energy
integral of the transmission function integrated over the bias window. The LB-DFT framework has become extremely
useful in a qualitative understanding of transport through, e.g., single molecules.

However, one should keep in mind that the non-interacting nature of electrons
in LB-DFT clearly is an approximation. Furthermore, LB-DFT uses ground-state
(equilibrium) DFT in a non-equilibrium situation (transport) which is formally
not justified.

A proper non-equilibrium DFT approach to transport could be time-dependent DFT
(TDDFT)\cite{RungeGross:84} where the steady-state is achieved in the
long-time limit of the time evolution of the system after switching on a DC
bias. Formally, the long-time limit of TDDFT leads to exchange-correlation
(xc) corrections to the bias
\cite{sa-1.2004,sa-2.2004,szvv.2005,kbe.2006,StefanucciKurthRubioGross:06,VignaleVentra:09}
which are absent in LB-DFT but are difficult to model. Within TDDFT, one can
hope to describe the (longitudinal) part of the electronic (steady)
current. For the additional description of heat (or energy) currents, the
formalism has been extended recently
\cite{EichVentraVignale:14,EichPrincipiVentraVignale:14,EichVentraVignale:17},
but applications have so far been restricted to non-interacting systems\cite{covito2018transient}.

Recently, an alternative DFT approach to transport in the steady state was suggested
\cite{stefanucci2015steady}. This so-called i-DFT formalism allows to compute the
steady-state density and electronic current (and thus the electrical conductance). Again,
just like in TDDFT, this is achieved via an xc contribution to the bias. Unlike in TDDFT,
however, xc functionals have been constructed for non-trivial model systems such as the
single impurity Anderson model (SIAM), both in the Kondo as well as in the Coulomb
blockade regime\cite{kurth2016nonequilibrium}. Also, in TDDFT
  the exact xc functional has memory dependence\cite{MaitraBurkeWoodward:02,Maitra:16,dittmann2018nonadiabatic,dittmann2019,Wijewardane2005,DAgosta2006b} whereas the i-DFT xc functionals only depends on the steady state
  values of the densities.  Somewhat unexpectedly, i-DFT can also be used
to compute many-body spectral functions both in \cite{JacobKurth:18,JacobStefanucciKurth:20} and
out of equilibrium \cite{kurth2019nonequilibrium}. 

By construction, i-DFT does not give access to the heat current (although the Seebeck
coefficient can be extracted\cite{sobrino2019steady}).
In the present work, we will close this gap and generalize i-DFT to
iq-DFT, a new formalism which allows to compute not only the (steady-state)
density and electrical current but also the heat current. 

The structure of the paper is as follows: in Sec.~\ref{formalism}, we prove the fundamental
theorem of the iq-DFT formalism (the one-to-one correspondence between densities and
potentials) and introduce the corresponding KS scheme. In Sec.~\ref{linres}, we derive the
linear response equations which allow expressing the linear transport coefficients
solely in terms of iq-DFT quantities. These coefficients include the thermal
conductance, for which up to now only the (formally incomplete) LB-DFT expression has been
available. In Sec.~\ref{siam}, we apply iq-DFT to the SIAM in
the linear response regime. From reverse engineering, we derive analytic parametrizations
for all iq-DFT quantities needed to describe the Anderson model in the Coulomb
blockade regime. In the Appendix, we give detailed derivations for the analytical
integrals needed in Sec.~\ref{siam}.

\section{Formalism}
\label{formalism}

We consider the typical setup for electronic transport which consists of a
central molecular junction ($C$) coupled to a left ($L$) and a right ($R$)
electrode. The electrodes are in (local) thermal equilibrium with temperatures
$T_{L/R}$ and chemical potentials $\mu_{L/R}$, respectively. The central
region $C$ is subject to an electrostatic potential $v(\mathbf{r})$ generated
by, e.g., the nuclei in the molecular junction and/or an external gate
potential which vanishes deep inside the electrodes. The system can be driven
out of equilibrium by a finite thermal gradient
$\Delta T=T_{L}-T_{R}$ and/or an external DC bias $V$ across the
junction. We assume that these perturbations in the long-time limit lead to
a steady-state electrical current ($I$) as well as energy and
  heat currents ($W$ and $Q$, respectively).

We aim to construct a formally exact density functional framework, which we
call iq-DFT, to describe such a steady state and reproduce these currents. 
To this end, we extend the recently proposed
DFT framework for steady-state transport \cite{stefanucci2015steady}, also
called i-DFT, which in principle captures the steady state
density $n(\mathbf{r})$ in the central region $C$, as well as the steady
current $I$ through it. By construction, in the linear-response regime, i-DFT
gives access to the (many-body) electrical conductance and can also describe
the Seebeck coefficient \cite{sobrino2019steady}. On the other hand, the
energy or heat currents are not guaranteed to be reproduced in i-DFT and
therefore also the thermal conductance is not captured. i-DFT is based on the
one-to-one map between the pair of ``densities'' ($n(\mathbf{r}),I$) and
the pair of potentials ($v(\mathbf{r}),V$), where the bias $V$ across region
$C$ is given as $V= \mu_L-\mu_R$.

In our new iq-DFT framework for the description of both charge {\em and}
thermal transport, we establish a one-to-one map between the three
``densities'' ($n(\mathbf{r}), I, Q$) and the three ``potentials''
($v(\mathbf{r}), V, \Psi$), where $\Psi=(T_L-T_R)/T$ is the normalized
thermal gradient and $T=(T_L+T_R)/2$ is the background temperature. In linear
response, iq-DFT gives access not only to the electrical conductance and
the Seebeck coefficient but also to the thermal conductance, see
Section \ref{linres}. 

In the following, we adopt the sign convention that currents flowing into the
central region are positive. Due to charge and energy conservation, the
steady-state electrical/energy current flowing in from the left lead is equal
to the steady-state electrical/energy current flowing out through the right
lead, i.e., $I\equiv I_{L}=-I_{R}$ (electrical current), $W\equiv W_{L}=-W_{R}$
(energy current) and  $Q\equiv Q_{L}=-IV-Q_{R}$ (heat current).

The foundation of iq-DFT rests on the following theorem which 
establishes the one-to-one correspondence between
the basic variables of the theory $(n(\mathbf{r}),I,Q)$ and the three
driving forces or potentials $(v(\mathbf{r}),V,\Psi)$. 

{\em Theorem}: 
For any finite temperature $T$ and fixed electrostatic potential in the leads,
there exists a one-to-one correspondence between the set of ``densities''
$(n(\mathbf{r}),I,Q)$ and the set of ``potentials'' $(v(\mathbf{r}),V,\Psi)$
in a (gate dependent) finite region of bias and thermal gradient around
$V=0$ and $\Psi=0$.

{\em Proof}: 
The existence of the invertible map can be proven by showing that the
determinant of the Jacobian 
\begin{gather}
  J=\text{det}\left.\left( {\begin{array}{ccc}
   \frac{\delta n(\mathbf{r})}{\delta v(\mathbf{r'})}  & \frac{\partial n(\mathbf{r})}{\partial V} & \frac{\partial n(\mathbf{r})}{\partial\Psi}  \\
   \frac{\delta I}{\delta v(\mathbf{r'})}  & \frac{\partial I}{\partial V} & \frac{\partial I}{\partial\Psi} \\
   \frac{\delta Q}{\delta v(\mathbf{r'})}   & \frac{\partial Q}{\partial V} & \frac{\partial Q}{\partial\Psi} \\
  \end{array} } \right)\right|_{\substack{V=0\\\Psi=0}},
   \label{matrix_J}
\end{gather}
is nonvanishing.
 
Since a change in the gate voltage can not produce a persistent current in
the linear regime we have
\begin{align}
\left.\frac{\delta I}{\delta v(\mathbf{r})}\right|_{\substack{V=0\\ \Psi=0}}=0, 
\left.\quad \frac{\delta Q}{\delta v(\mathbf{r})}\right|_{\substack{V=0\\\ \Psi=0}}=0;
\label{eq:vanishing_derivatives}
\end{align}
and
therefore we can write Eq.~(\ref{matrix_J}) as
\be
J = \text{det}(\chi(\mathbf{r},\mathbf{r'}))\text{det}(\mathbf{L})
=\text{det}(\chi(\mathbf{r},\mathbf{r'}))TG\kappa,
\label{eq:det_J}
\ee
where $\chi(\mathbf{r},\mathbf{r'})=\frac{\delta n(\mathbf{r})}{\delta v(\mathbf{r'})}\Big|_{\substack{ V=0\\\ \Psi=0}}$
is the static equilibrium density response function and $G$ and $\kappa$ are
the electrical and thermal conductances. These are derived from the linear response
relationship between the currents $( I, Q)$ which, to first order,
result from application of the potentials $(V, \Psi)$ 
\begin{gather}
  \left( {\begin{array}{cc}
   I \\  Q
  \end{array} } \right)
  = 
  \textbf{L}
    \left( {\begin{array}{cc}
    V\\  \Psi
  \end{array} } \right)
  = 
  \left( \begin{array}{cc}
    L_{11} & L_{12} \\
    L_{21} & L_{22} \end{array} \right)
    \left( {\begin{array}{cc}
   V\\  \Psi
  \end{array} } \right)
   \label{relat_vary}
\end{gather}
with $L_{21}=L_{12}$ from Onsager's relation\cite{onsager1931reciprocal}. 
The conductance matrix $\textbf{L}$ can be expressed in terms of the
transport coefficients as \cite{callen1948application}
\begin{gather}
   \mathbf{L}= 
 \left( {\begin{array}{cc}
   G  & -TG S \\
   -TG S  & T\kappa +T^{2}G S^{2}\\
  \end{array} } \right)
  \label{matrix_L}
\end{gather}
where $S$ is the Seebeck coefficent. Equivalently, we can 
use 
Eq.~(\ref{matrix_L}) to express the transport coefficients in terms of the
matrix elements $L_{ij}$ as 
\begin{subequations}
\begin{align}
G=&\left.\frac{\partial I}{\partial V}\right|_{\substack{V=0\\\ \Psi=0}}=L_{11},
\label{eq:G}
\end{align}
\begin{align}
S=&\left.\frac{\partial V}{\partial \Delta T}\right|_{\substack{I=0\\\ Q=0}}=-\frac{1}{T} \frac{L_{12}}{L_{11}},
\end{align}
\label{th_coeff_S}
\begin{align}
\kappa = &-\left.\frac{\partial Q}{\partial \Delta T}\right|_{\substack{I=0\\\ Q=0}}=\frac{1}{T}\left(L_{22}-\frac{L_{12}^{2}}{L_{11}}\right).
\label{k_th_Q}
\end{align}
\label{eqs:th_coeffs}
\end{subequations}
It has already been shown \cite{stefanucci2015steady} that, for any finite
temperature $T$, we have $\text{det}(\chi(\mathbf{r},\mathbf{r'}))<0$
\cite{stefanucci2015steady} and $G>0$ \cite{stefanucci2015steady,bsw.2006}.
Therefore, in order to complete the proof of the theorem, it remains to be
shown that $\kappa>0$.

This step can be done by applying Onsager's original arguments.
In the steady-state the time-derivative of the entropy $\mathcal{S}$ in the
central molecular region equals the sum of the entropy currents $Q_\alpha/T$, $\alpha=L,R$,
from the leads
\bea
\dot{\mathcal{S}} &=& 
-\left(\frac{ Q_L}{T_L} + \frac{ Q_R}{T_R} \right) \nonumber \\
&=& -\frac{1}{T} \left(  Q_L+ Q_R + \frac{ \Psi}{2}
( Q_L- Q_R) \right)
\label{entropy_rate}
\eea
where the minus sign follows from our
convention that currents flowing into the central region $C$ are
positive. In the last
step, we expanded to linear order for small temperature gradients $\Psi$. 
Using the relation 
\begin{align}
 W_{\alpha}= Q_{\alpha}+\mu_{\alpha} I_{\alpha}.
\label{eq:W_to_Q}
\end{align}
between energy and heat currents, and using that in the steady-state
we have $ I_L=-I_R= I$ and $ W_L=- W_R =  W$, we arrive at 
\bea
\dot{\mathcal{S}} &=& \frac{1}{T}\left( I  V + Q  \Psi \right) \nonumber\\
&=& \frac{1}{T}\left(L_{11} V^{2} +2L_{12} V \Psi+L_{22} \Psi^{2} \right) 
\label{eq:S_dot}
\eea
where we used Eq.~(\ref{relat_vary}) in the last step.

From the second law of the thermodynamics, we know that
$\dot{\mathcal{S}}\ge 0$ where the equality sign holds at equilibrium.
Therefore, the equilibrium state of the system corresponds to the local minimum
of the rate of production of entropy. As a consequence, the determinant of the Hessian
matrix of $\dot{\mathcal{S}}$ has to be positive around $V=0,\Psi=0$
\begin{align}
  \mathbf{H}_{\dot{S}}&=\left.\frac{\partial^{2} \dot{\mathcal{S}}}{\partial V^{2}}
  \right|_{\substack{V=0\\ \Psi=0}}
  \left.\frac{\partial^{2}\dot{\mathcal{S}}}{\partial\Psi^{2}}
  \right|_{\substack{V=0\\ \Psi=0}}
  -\left.\frac{\partial^{2}\dot{\mathcal{S}}}{\partial V\partial\Psi}
  \right|_{\substack{V=0\\\Psi=0}}^{2}\nonumber\\
&= \frac{4}{T^{2}}\left(L_{11}L_{22}-L_{12}^{2}\right)=4T^{-1}G\kappa>0.
\label{eq:k_positive}
\end{align}
From Eq.~(\ref{eq:k_positive}), the positiveness of the
thermal conductance $\kappa>0$ directly follows
which completes the proof of the one-to-one map.

An equivalent formulation stems from considering as third basic variable the energy 
current $W$ instead of $Q$. The theory thus leads to a one-to-one correspondence 
between $(n(\mathbf{r}),I,W)$ and the trio of potentials
$(v(\mathbf{r}),V-\mu_{L}\Psi,\Psi)$. The two formulations are related through Eq.~(\ref{eq:W_to_Q}).\cite{sierra2015nonlinear}

\subsection{Kohn-Sham equations of iq-DFT}

The iq-DFT theorem holds for any form of the interaction, in
particular also for the noninteracting case. In order to establish the
Kohn-Sham (KS) scheme, we make the usual assumption of non-interacting
representability, i.e., that there exists a unique trio of potentials
$(v_{s}(\mathbf{r}), V_{s}, \Psi_{s})$ for a non-interacting system, the Kohn-Sham
system, which exactly reproduces the densities $(n(\mathbf{r}),I,Q)$ of the
interacting system with potentials $(v(\mathbf{r}), V, \Psi)$.
Following the standard KS procedure, the xc potentials of the iq-DFT
framework are then defined as 
\begin{subequations}
\begin{align}
v_{\rm Hxc}[n,I,Q](\mathbf{r}) &=v_{s}[n,I,Q](\mathbf{r}) -v[n,I,Q](\mathbf{r}) ,\\
V_{\rm xc}[n,I,Q] &=V_{s}[n,I,Q] -V[n,I,Q], \\
\Psi_{\rm xc}[n,I,Q]&=\Psi_{s}[n,I,Q] -\Psi[n,I,Q].
\end{align}
\label{eq:XC_potentials}
\end{subequations}

The self-consistent coupled KS equations for the densities read
($\int\equiv\int_{-\infty}^{\infty} \frac{d\omega}{2\pi}$ in the following)
\begin{subequations}
\begin{align}
&n(\mathbf{r})=2\sum_{\alpha=L,R}\int f(\frac{\omega-\mu_{s,\alpha}}{T_{s,\alpha}})A_{s,\alpha}(\mathbf{r},\omega)\label{eq:KS_densities_n},\\
&I=2\sum_{\alpha=L,R}\int f(\frac{\omega-\mu_{s,\alpha}}{T_{s,\alpha}})s_{\alpha}\mathcal{T}_{s}(\omega)\label{eq:KS_densities_I},\\
&Q=2\sum_{\alpha=L,R}\int f(\frac{\omega-\mu_{s,\alpha}}{T_{s,\alpha}})s_{\alpha}(\omega-\mu_{s,L}) \mathcal{T}_{s}(\omega),\label{eq:KS_densities_Q}
\end{align}
\label{eq:KS_densities}
\end{subequations}
where $f(x)=[1+\exp(x)]^{-1}$
is the Fermi function, $\mu_{s,\alpha}=\mu+V_{s,\alpha}$,
$T_{s,\alpha}=T(1+s_{\alpha}\Psi_s/2)$ and $s_{L/R}= \pm 1$. We also defined the
partial spectral function
$A_{s,\alpha}(\mathbf{r},\omega)=\bra{\mathbf{r}}\mathcal{G}(\omega)
\Gamma_{\alpha}(\omega)\mathcal{G}^{\dagger}(\omega)\ket{\mathbf{r}}$, with
$\mathcal{G}(\omega)$ and $\Gamma_{\alpha}(\omega)$ the KS Green's function and
broadening matrices, respectively, and the KS transmission function
$\mathcal{T}_{s}(\omega)=\text{Tr}\left\{\mathcal{G}(\omega)
\Gamma_{L}(\omega)\mathcal{G}^{\dagger}(\omega)\Gamma_{R}(\omega)\right\}$.
Finally, the energy current follows directly from  Eqs.~(\ref{eq:W_to_Q}) and
(\ref{eq:KS_densities})
\begin{align}
W=2&\sum_{\alpha=L,R}\int f(\frac{\omega-\mu_{s,\alpha}}{T_{s,\alpha}})s_{\alpha}\omega\mathcal{T}_{s}(\omega).
\end{align}

Eqs.~(\ref{eq:KS_densities_n}) and (\ref{eq:KS_densities_I}) have the same
structure as the KS equations of the original i-DFT formulation, except that
in the present formalism the thermal gradient along the central region is not
a parameter anymore but a basic potential which depends on the densities of
the system. Therefore, the only possible parametric temperature dependence in
the approximations for the functionals Eqs.~(\ref{eq:XC_potentials}) is
through the average temperature $T$.

\section{Linear Response}
\label{linres}

In this section, we develop the linear response formalism for iq-DFT which
leads to expressions for the linear transport coefficients $G$, $S$, and
$\kappa$ purely in terms of quantities accessible by the theory.

The linear relationship for small variations of the basic densities around
zero currents follows Eq.~(\ref{relat_vary}). The same current variations can
be expressed in terms of the KS system

\begin{gather}
  \left( {\begin{array}{cc}
   I \\
    Q
  \end{array} } \right)
  = 
\textbf{L}_{s}
    \left( {\begin{array}{cc}
  V+  V_{\rm xc}\\
  \Psi+\Psi_{\rm xc}
  \end{array} } \right),
  \label{KS_changes}
\end{gather}
where we have used the definition of the KS potentials
Eqs.~(\ref{eq:XC_potentials}) and that $ I_{s}= I$ and
$ Q_{s}= Q$ by the KS construction. In the linear response regime,
the changes in the xc potentials can be written as
\begin{gather}
  \left( {\begin{array}{cc}
  V_{\rm xc}\\
 \Psi_{\rm xc}
  \end{array} } \right)
  = 
\textbf{F}_{\rm xc}
    \left( {\begin{array}{cc}
   I\\
  Q
  \end{array} } \right),
  \label{xc_changes}
\end{gather}
with the matrix of xc derivatives $\mathbf{F}_{\rm xc}$ defined by
\begin{gather}
   \mathbf{F}_{\rm xc}=\left.
  \left( {\begin{array}{cc}
\frac{\partial V_{\rm xc}}{\partial I}  &  \frac{\partial V_{\rm xc}}{\partial Q} \\
       \frac{\partial \Psi_{\rm xc}}{\partial I}  &  \frac{\partial \Psi_{\rm xc}}{\partial Q} 
  \end{array} } \right)  \right |_{\substack{I=0\\\ Q=0}}.
  \label{matrix_F_xc}
\end{gather}
Combining Eqs.~(\ref{relat_vary}), (\ref{KS_changes}), and (\ref{matrix_F_xc}), and using the fact that $V$ and $\Psi$ are arbitrary, 
we arrive at the Dyson equation

\begin{gather}
\textbf{L}
  = 
\textbf{L}_{s}+\textbf{L}_{s}\textbf{F}_{\rm xc}\textbf{L},
  \label{Dyson}
\end{gather}

or, equivalently, 
\begin{align}
\mathbf{F}_{\rm xc}=\mathbf{L}_{s}^{-1}-\mathbf{L}^{-1}=\mathbf{R}_{s}-\mathbf{R}.
\label{Dyson_XC}
\end{align}
Here, $\mathbf{L}$ and $\mathbf{L}_{s}$ are the interacting and KS conductance
matrices where each element is evaluated at $(V=0,\Psi=0)$ and
$(V_s=0,\Psi_s=0)$, respectively. Similarly, $\mathbf{R}=\mathbf{L}^{-1}$ and
$\mathbf{R}_{s}=\mathbf{L}_{s}^{-1}$ are the interacting and KS resistance matrices
where each element is evaluated at $(I=0,Q=0)$. As a consequence of the
Onsager's relations between the cross terms in the conductance matrices, from Eq.~(\ref{Dyson_XC}) it follows
\begin{align}
  \left.\frac{\partial V_{\rm xc}}{\partial Q}\right|_{\substack{I=0\\\ Q=0}}=\left.\frac{\partial
    \Psi_{\rm xc}}{\partial I}\right|_{\substack{I=0\\\ Q=0}}.
\label{onsager_xc_derivs}
\end{align}
We can express the $\mathbf{F}_{\rm xc}$ elements as function of the linear
transport coefficients making use of Eqs.~(\ref{matrix_L}) and
(\ref{Dyson_XC}) for the interacting and the KS system
\begin{subequations}
 \begin{align}
 \frac{\partial V_{\rm xc}}{\partial I}\Bigr|_{\substack{I=0\\ Q=0}}=&\frac{1}{G_s}+T\frac{S_{s}^{2}}{\kappa_s}-\frac{1}{G}-T\frac{S^{2}}{\kappa},\\
 \frac{\partial \Psi_{\rm xc}}{\partial I}\Bigr|_{\substack{I=0\\ Q=0}}=& \frac{\partial V_{\rm xc}}{\partial Q}\Bigr|_{\substack{I=0\\ Q=0}}=\frac{S_s}{\kappa_s}-\frac{S}{\kappa},\\
 \frac{\partial \Psi_{\rm xc}}{\partial Q}\Bigr|_{\substack{I=0\\ Q=0}}=&\frac{1}{T\kappa_s}-\frac{1}{T\kappa}.
 \end{align}
  \label{eq:xc_currents_to_zero}
\end{subequations} 
These equations can be inverted to express the transport coefficients as 
\begin{subequations}
 \begin{align}
 \kappa= &\frac{\kappa_{s}}{1-T\frac{\partial \Psi_{\rm xc}}{\partial Q}\Bigr|_{\substack{I=0\\ Q=0}}\kappa_{s}},\label{eq:heat_conductance_R}\\
 S=&\frac{S_{s}-\kappa_{s}\frac{\partial V_{\rm xc}}{\partial Q}\Bigr|_{\substack{I=0\\ Q=0}}}{1-T\frac{\partial \Psi_{\rm xc}}{\partial Q}\Bigr|_{\substack{I=0\\ Q=0}}\kappa_{s}},\label{eq:Seebeck_R}\\
G=&\frac{G_{s}}{1-\left(\frac{\partial V_{\rm xc}}{\partial I}\Bigr|_{\substack{I=0\\ Q=0}}+\frac{TS^{2}}{\kappa}-\frac{TS_{s}^{2}}{\kappa_s}\right)G_{s}}.\label{eq:el_conductance_R}
 \end{align}
 \label{eq:R_thermal_coeff} 
\end{subequations}
Eqs.~(\ref{eq:R_thermal_coeff}) are exact expressions for the interacting
(linear) transport coefficients in any molecular transport setup. They express
the many-body transport coefficients in terms of quantities which are fully
accessible within iq-DFT, i.e, the xc derivatives evaluated at $(I=0, Q=0)$
and the KS transport coefficients. The transport coefficients in iq-DFT
exhibit increasing complexity: while the thermal conductance $\kappa$
(Eq.~(\ref{eq:heat_conductance_R})) only depends on the KS thermal conductance
$\kappa_{s}$ and $\frac{\partial \Psi_{\rm xc}}{\partial Q}$,
the Seebeck coefficient depends on its KS contribution $S_{s}$, $\kappa_s$ as
well as the two xc derivatives, $\frac{\partial V_{\rm xc}}{\partial Q}$ and
$\frac{\partial \Psi_{\rm xc}}{\partial Q}$. Finally, the electrical conductance
depends on the three KS coefficients ($\kappa_{s},S_{s},G_{s}$) and the three
xc derivatives through $S$ and $\kappa$.

Using Eqs.~(\ref{eq:R_thermal_coeff}) for the iq-DFT transport coefficients, we
now briefly discuss the relation of iq-DFT to other DFT-based frameworks for
the description of steady-state transport. At first, we consider the simplest
approximation which completely neglects the xc contributions to the
transport coefficients, i.e., setting $V_{\rm xc}\approx 0$ and
$\Psi_{\rm xc}\approx 0$. Then all linear transport coefficients reduce to the
corresponding KS coefficients, i.e., we recover the standard LB-DFT
approach. At the next level, we consider the relation to the original i-DFT
formalism which is designed to give the exact electrical steady current. 
The i-DFT expression for the electrical conductance
\be
G = \frac{G_s}{1-\frac{\partial V_{\rm xc}^{i-DFT}}{\partial I}\Bigr|_{\substack{I=0}}
  G_s}
\label{cond_idft}
\ee
is exact, just as the corresponding iq-DFT expression
(\ref{eq:el_conductance_R}). Thus, we can establish the exact relation
\begin{align}
  \frac{\partial V_{\rm xc}^{i-DFT}}{\partial I}\Bigr|_{\substack{I=0}} =
  \frac{\partial V_{\rm xc}^{iq-DFT}}{\partial I}\Bigr|_{\substack{I=0\\ Q=0}}
  +\frac{TS^{2}}{\kappa}-\frac{TS_{s}^{2}}{\kappa_s}
\end{align}
for the current derivatives at $I=0$ of the xc bias in i-DFT and iq-DFT.
In the original i-DFT framework, the Seebeck coefficient as well as the thermal
conductance are given by their KS counterparts. In iq-DFT, this corresponds to
the approximation of setting $\Psi_{\rm xc}\approx 0$ and approximating the
xc bias as a functional independent of the heat current, i.e.,
$V_{\rm xc}[n,I,Q] \approx V_{\rm xc}[n,I]$. In earlier work
\cite{sobrino2019steady}, we have extended the original i-DFT formalism to not
only give the many-body electrical conductance but also the many-body Seebeck
coefficient, while for the thermal conductance one still has $\kappa=\kappa_s$.
In iq-DFT, this corresponds to the approximation
$\Psi_{\rm xc}[n,I,Q] \approx \Psi_{\rm xc}[n,I]$, independent of $Q$ for general
$V_{\rm xc}[n,I,Q]$. Then we find
$S=S_{s}-\kappa_{s}\frac{\partial V_{\rm xc}}{\partial Q}|{\substack{I=0\\ Q=0}}=
S_{s}+\frac{\partial V_{\rm xc}}{\partial \Delta T}|{\substack{I=0\\ Q=0}}=S_{s}+S_{\rm xc}$, as in
Refs.~\onlinecite{sobrino2019steady,yang2016density}. This approximation then also implies
a finite correction (over pure i-DFT) for the electrical conductance
$G=G_{s}[1-\left(\frac{\partial V_{\rm xc}}{\partial I}|_{\substack{I=0\\ Q=0}}+
TS_{\rm xc}(S_{\rm xc}-2S_s)/\kappa_s\right)G_{s}]^{-1}$.
  
In order to calculate the interacting transport coefficients from
Eq.~(\ref{eq:R_thermal_coeff}), one first needs to evaluate the KS
coefficients, and consequently, an approximation for the functional
$v_{\rm Hxc}[n]$ is required where the dependence of $v_{\rm Hxc}$ on $I$ and $Q$
can be neglected if we work in the linear response regime.
In order to gain some first insight into the possible approximations
for the iq-DFT functionals, in the following Section we will discuss an 
application of iq-DFT formalism to a particular model system in the linear
response regime.

\section{Application to the single impurity Anderson Model}
\label{siam}

In this section we apply our iq-DFT framework to the SIAM. Due to its simplicity and evident physical
interpretation, this model is ideally suited as a first system to explore the new formalism and has
been used in many previous works \cite{kurth2019nonequilibrium,yang2016density,alomar2016coulomb}, both
within and outside any DFT setting. The SIAM describes a single interacting impurity (quantum dot)
coupled to non-interacting left (L) and right (R) leads.
The Hamiltonian of the system reads
\begin{align}
  \hat{H} = &\sum_{\sigma}v \hat{n}_{\sigma} + U\hat{n}_{\uparrow}\hat{n}_{\downarrow}+ \sum_{\alpha k \sigma}\varepsilon_{\alpha k \sigma} \hat{c}^{\dagger}_{\alpha k \sigma} \hat{c}_{\alpha k \sigma}\nonumber\\
& + \sum_{k \alpha \sigma}\left( t_{\alpha k}\hat{c}^{\dagger}_{\alpha k \sigma}\hat{d}_{\sigma} + H.c.\right).
\label{eq:SIAM_H}
\end{align}
The first two terms in Eq.~(\ref{eq:SIAM_H}) describe the single impurity, where $v$ is the
on-site energy of the dot and $U$ is the Coulomb interaction.
$\hat{c}_{\alpha k \sigma}^{\dagger}$ and $\hat{d}_{\sigma}^{\dagger}$ are the creation operators for
electrons with spin $\sigma$ ($\sigma=\uparrow,\downarrow$) in lead $\alpha$ and on the dot, respectively.
$\hat{n}_{\sigma}=\hat{d}^{\dagger}_{\sigma}\hat{d}_{\sigma}$ and
$\hat{n}=\hat{n}_{\uparrow}+\hat{n}_{\downarrow}$ are the operators for the spin density and for
the total density of electrons on the dot. The third and last term account for the single
particle eigenstates of the isolated leads as well as for the tunnelling between the dot and
the leads with couplings $\Gamma_{\alpha}(\omega) := 2 \pi \sum_k |t_{\alpha k}|^2
\delta(\omega -\varepsilon_{k \alpha})$. 
We consider featureless electronic leads described by frequency-independent couplings
$\Gamma_{\alpha}(\omega)=\gamma_{\alpha}$ (with $\alpha=L,R$), i.e., we work in the wide band limit (WBL).
For simplicity, we choose symmetric coupling of the leads, i.e., $\gamma_{L}=\gamma_{R}=\gamma/2$.
In the present Section we are mostly concerned with application of the theory to the linear response
regime, but for derivation purposes we keep a finite symmetric thermal gradient and a finite symmetric
DC bias between the leads , i.e., $T_{\alpha}=T(1+s_{\alpha}\frac{\Psi}{2})$ and $V_{\alpha}=s_{\alpha}\frac{V}{2}$
with $s_{L/R}= \pm 1$ where we choose $\mu=0$.    

\subsection{Reverse engineering from a many body model}
In order to apply our iq-DFT formalism to the SIAM, we need approximations for all the xc potentials of
the formalism. Since here we are concerned with the linear response regime only, we actually need to
construct parametrizations for the derivatives of the xc potentials (at zero currents) appearing
in Eqs.~(\ref{eq:R_thermal_coeff}).

This can be achieved through a reverse engineering process. First, we express the interacting
density on and currents through the dot in terms of the many-body spectral function
$A(\omega)$ \cite{MeirWingreen:92,CostiZlatic:10}: 
\begin{subequations}
\begin{align}
  n=&\sum_{\alpha=L,R}\int f\left(\frac{\omega-s_{\alpha}\frac{V}{2}}{T 
    \left(1+s_{\alpha}\frac{\Psi}{2}\right)}\right)A (\omega),\label{eq:int_eqs_a}\\
  I=&\frac{\gamma}{2}\sum_{\alpha=L,R}\int s_{\alpha}f\left(\frac{\omega-s_{\alpha}\frac{V}{2}}{T
    \left(1+s_{\alpha}\frac{\Psi}{2}\right)}\right)A (\omega),\\
  Q=&\frac{\gamma}{2}\sum_{\alpha=L,R}\int s_{\alpha}f\left(\frac{\omega-s_{\alpha}\frac{V}{2}}{T
    \left(1+s_{\alpha}\frac{\Psi}{2}\right)}\right)(\omega-\frac{V}{2}) A(\omega).\label{eq:densities_Q}
\end{align}
\label{eq:int_eqs}
\end{subequations}
 In order to proceed with the reverse engineering, we consider the
following model for the many-body spectral function \cite{kurth2017transport} which correctly
describes the impurity coupled to the leads in the parameter range $T/\gamma>1$
\begin{align}
  A(\omega)=\left(1-\frac{n}{2}\right)\frac{\gamma}{(\omega-v)^{2} +\frac{\gamma^2}{4}}
  +\frac{n}{2}\frac{\gamma}{(\omega-v-U)^{2}+\frac{\gamma^2}{4}}.
\label{eq:model_spectral_function_SIAM}
\end{align}
In the following, we denote the spectral function of Eq.~(\ref{eq:model_spectral_function_SIAM})
as many-body model (MBM). This model can be derived from the equations of motion
technique \cite{jauho-book}. However, it may also be understood more intuitively by
calculating the exact spectral function of the single site model (SSM)\cite{stefanucci2011towards}, i.e., the limit of the
uncontacted impurity. Broadening the delta peaks of the SSM spectral function to Lorentzian
peaks with width given by the coupling strength $\gamma$, one obtains
Eq.~(\ref{eq:model_spectral_function_SIAM}). The validity of this model is formally limited to
temperatures larger than any other energy scale of the system. In particular, Eq.~(\ref{eq:model_spectral_function_SIAM}) correctly captures
  Coulomb blockade physics, but not the Kondo regime.

For the reverse engineering, we also need the densities and currents expressed through the
KS equations. These can be obtained from Eqs.~(\ref{eq:int_eqs}) by replacing the basic
potentials by their non-interacting versions, i.e., $v\to v_s$, $V\to V_s$ and $\Psi \to \Psi_s$,
and replacing $A(\omega)\to A_{s}(\omega)=\gamma/((\omega-v_s)^{2}+\gamma^2/4)$. The resulting
integrals can be evaluated analytically (see Appendix) and the basic variables of the theory
can then be expressed as 
\begin{subequations}
\begin{align}
n=&1 -\frac{1}{\pi}\left(\operatorname{Im}{\left[\psi\left(z_{s}^{R}\right)\right]}+\operatorname{Im}{\left[\psi\left(z_{s}^{L}\right)\right]}\right),\\
I=&\frac{\gamma}{2\pi}\left(\operatorname{Im}{\left[\psi\left(z_{s}^{R}\right)\right]}-\operatorname{Im}{\left[\psi\left(z_{s}^{L}\right)\right]}\right),\\
Q=&\frac{\gamma^{2}}{2\pi}\left(\operatorname{Re}{\left[\psi\left(z_{s}^{L}\right)\right]}-\operatorname{Re}{\left[\psi\left(z_{s}^{R}\right)\right]}\right)\nonumber\\
& +\frac{\gamma^{2}}{2\pi}\log{\left(\frac{1+\Psi_{s}/2}{1-\Psi_{s}/2}\right)}+\left(v_{s}-\frac{V_{s}}{2}\right)I,
\end{align}
\label{eq:digamma_eqs}
\end{subequations}
where $z_{s}^{L/R}=\frac{1}{2}+\frac{\gamma/2+\mathrm{i}(v_{s}\mp V_{s}/2)}{2\pi T(1\pm\Psi_{s}/2)}$
and $\psi(z)=\frac{d\log{(\Gamma(z))}}{dz}$ is the digamma function with general complex argument
$z$, and $\Gamma(z)$ is the gamma function.\cite{AbramowitzStegun:65}
 Also for our model many-body spectral function
(\ref{eq:model_spectral_function_SIAM}), all integrals in Eqs.~(\ref{eq:int_eqs}) can be
evaluated analytically using the same integrals of the Appendix but here we refrain
from showing the resulting expressions explicitly.

Taking the derivatives in Eqs.~(\ref{eq:digamma_eqs}) with respect to the related KS potentials, 
we can derive in an exact way the matrix elements $L_{ij}^{s}$ of the matrix ${\mathbf L}_s$
(see Eq.~(\ref{KS_changes})) as
\be
L_{ij}^{s}(v_{s})=M_{ij}(v_{s})
\label{eq:Lijs}
\ee
where we have used the $M_{ij}$ coefficients derived in Eqs.~(\ref{eq:Mij}) and made explicit the
dependence on the KS potential $v_s$.

Similarly, also for the many-body model (Eq.~(\ref{eq:model_spectral_function_SIAM}) inserted
into Eqs.~(\ref{eq:int_eqs})), we can derive the corresponding matrix elements of the
interacting response matrix ${\mathbf{L}}$ by taking the corresponding derivatives. These
matrix elements then read
\be
L_{ij}(v) = \left(1-\frac{n}{2}\right)M_{ij}(v)+\frac{n}{2}M_{ij}(v+U).
\label{eq:Gij}
\ee
Combining Eqs.~(\ref{eqs:th_coeffs}) and Eqs.~(\ref{eq:xc_currents_to_zero}) we arrive at 
\begin{subequations}
  \begin{align}
    \frac{\partial V_{\rm xc}}{\partial I}\Bigr|_{\substack{I=0\\
    Q=0}}=&\frac{1}{L_{11}^{s}(v_s)}
-\frac{1}{L_{11}(v)}+\frac{L_{12}^{s}(v_s)^2}{L_{11}^{s}(v_s)}\frac{1}{\det(\mathbf{L_s}(v_s))}\nonumber\\
&-\frac{L_{12}(v)^{2}}{L_{11}(v)}\frac{1}{\det(\mathbf{L}(v))},\\
\frac{\partial \Psi_{\rm xc}}{\partial I}\Bigr|_{\substack{I=0\\ Q=0}}=&
\frac{\partial V_{\rm xc}}{\partial Q}\Bigr|_{\substack{I=0\\ Q=0}}=\frac{L_{12}^{s}(v_s)}{\det(\mathbf{L_s}(v_s))}-\frac{L_{12}(v)}{\det(\mathbf{L}(v))},\\ 
\frac{\partial \Psi_{\rm xc}}{\partial Q}\Bigr|_{\substack{I=0\\ Q=0}}=&\frac{L_{11}^{s}(v_s)}{\det(\mathbf{L_s}(v_s))}-\frac{L_{11}(v)}{\det(\mathbf{L}(v))},
  \end{align}
  \label{eq:XC_derivs_G_ijs} 
\end{subequations}
 Eqs.~(\ref{eq:XC_derivs_G_ijs}) together with Eqs.~(\ref{eq:Gij}) provide the
analytical parametrizations of the xc derivatives in terms of both $v_s$ and $v$. 
\begin{figure}[t!]
  \includegraphics[width=0.98\columnwidth]{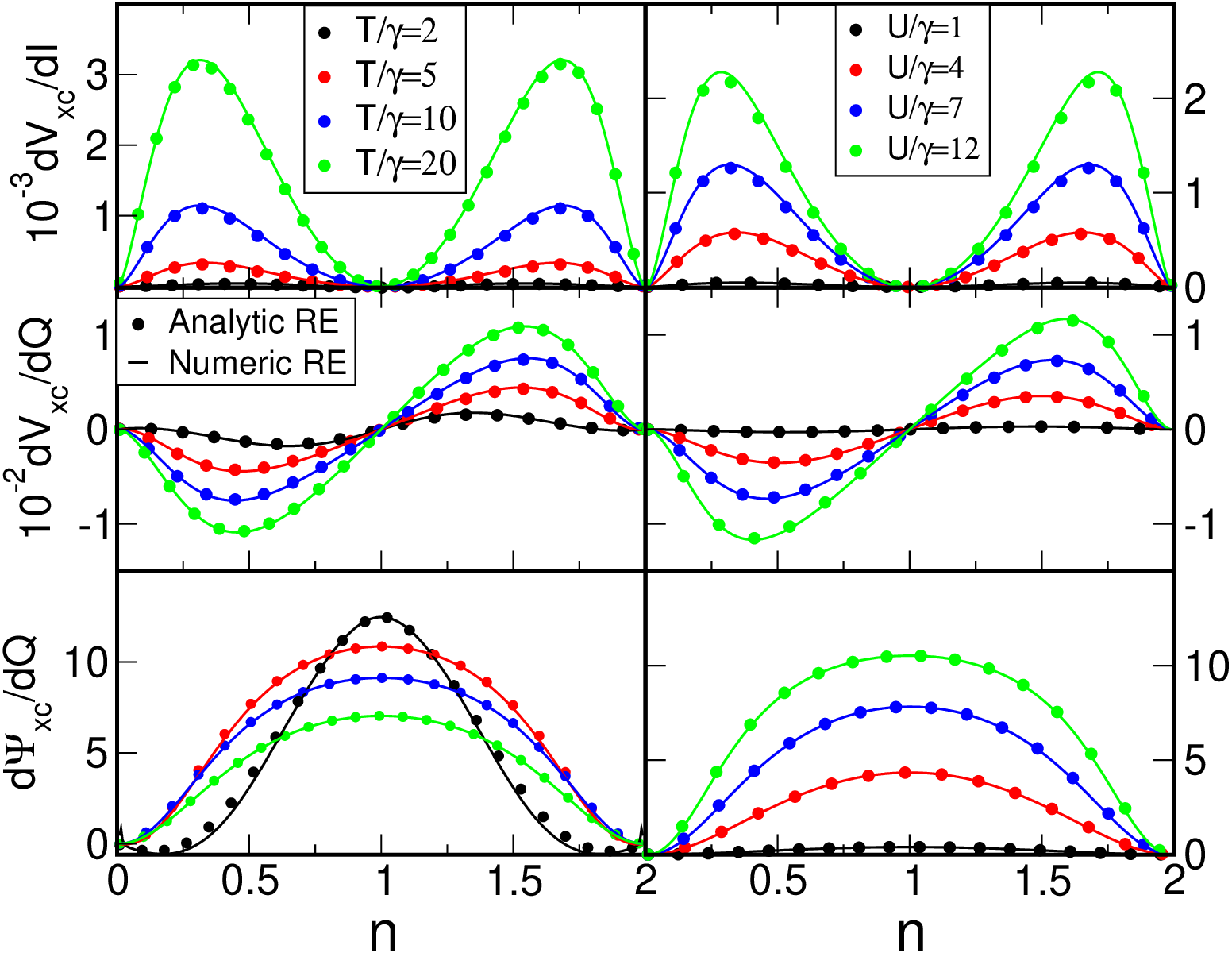}
  \caption{Comparison between analytical and numerical reverse engineered xc derivatives as function of the density. The left column corresponds to $U/\gamma=8$ and the right one to $T/\gamma=12$. For the analytic result the relation between the gates and the density from Eqs.~(\ref{eq:v_s_and_v_of_n}) has been used, while for the numeric inversion this relation directly follows Eq.~(\ref{eq:digamma_eqs}a). The xc derivatives are obtained in both approaches using Eqs.~(\ref{eq:Gij}) and (\ref{eq:XC_derivs_G_ijs}).}
  \label{fig:XC_derivatives}
\end{figure}
Instead, the
dependence of the xc derivatives on the density can be obtained by (i) replacing $v$ in the argument
of the many-body coefficients $L_{ij}$ by $v(n)$, the inverse of the density-potential relationship
of Eq.~(\ref{eq:int_eqs_a}) (at $V=0$ and $\Psi=0$) and, similarly, (ii) by using $v_s(n)$ as
arguments in the KS coefficients $L_{ij}^s$ which can be obtained by inverting the corresponding
KS expression $n(v_s)$ for the density (at $V_s=0$ and $\Psi_s=0$). These inverse functions
can easily be obtained numerically and, by construction, the resulting density functionals for the
xc derivatives then give exactly the same linear response transport coefficients (in a DFT
framework) as the many-body model. Nevertheless, here we are interested in finding an analytical
parametrization for the xc derivatives in terms of the density and therefore an approximation for
the density-gate relationship is required. 

Following ideas from previous works 
\cite{stefanucci2011towards,sobrino2020exchange}, we can refer to the 
SSM which describes a single (interacting or non-interacting) site not connected to leads but in
contact with a heat and particle bath. The exact density-gate relations for the non-interacting
and interacting SSM read
\begin{subequations}
 \begin{align}
v_{s}(n) = &T\log{\left(\frac{2}{n} - 1\right)}\label{eq:vs},\\
v(n) = &-U-T\log{\left(\frac{\delta n+\sqrt{\delta n^{2}+e^{- \frac{U}{T}}(1-\delta n^{2})}}{1-\delta n}\right)}\label{eq:v},
 \end{align}
 \label{eq:v_s_and_v_of_n} 
\end{subequations}
where $\delta n=n-1$. The SSM may be viewed as the limiting case of a SIAM weakly coupled to leads
and the Eqs.~(\ref{eq:v_s_and_v_of_n}) become more accurate as the ratio $T/\gamma$
increases\cite{sobrino2019steady}. Insertion of Eqs.~(\ref{eq:v_s_and_v_of_n}) into
Eqs.~(\ref{eq:XC_derivs_G_ijs}) defines our fully analytical parametrizations of the derivatives of the
xc potentials of iq-DFT in the linear esponse regime. These functionals provide a measure of the
correction required over the KS system to accurately describe the linear response properties of the
many-body model. From Fig.~\ref{fig:XC_derivatives} it is evident that the xc corrections become
larger with increasing temperature $T$ or interaction strength $U$. In the left column of
Fig.~\ref{fig:XC_derivatives}, the xc derivatives are calculated at $U/\gamma=8$ for different
temperatures $T$ while in the right column $T$ is fixed to $T/\gamma=12$ and the xc
derivatives are obtained for different interactions $U$. Our analytical parametrization is compared
with the numerically exact inversion from the MBM approach. As expected, we find excellent agreement between the approaches, in particular for temperatures $T$ larger than the other energies of the problem, i.e., the Coulomb interaction $U$ and the coupling to the leads $\gamma$.

\begin{figure}
  \includegraphics[width=0.9\columnwidth]{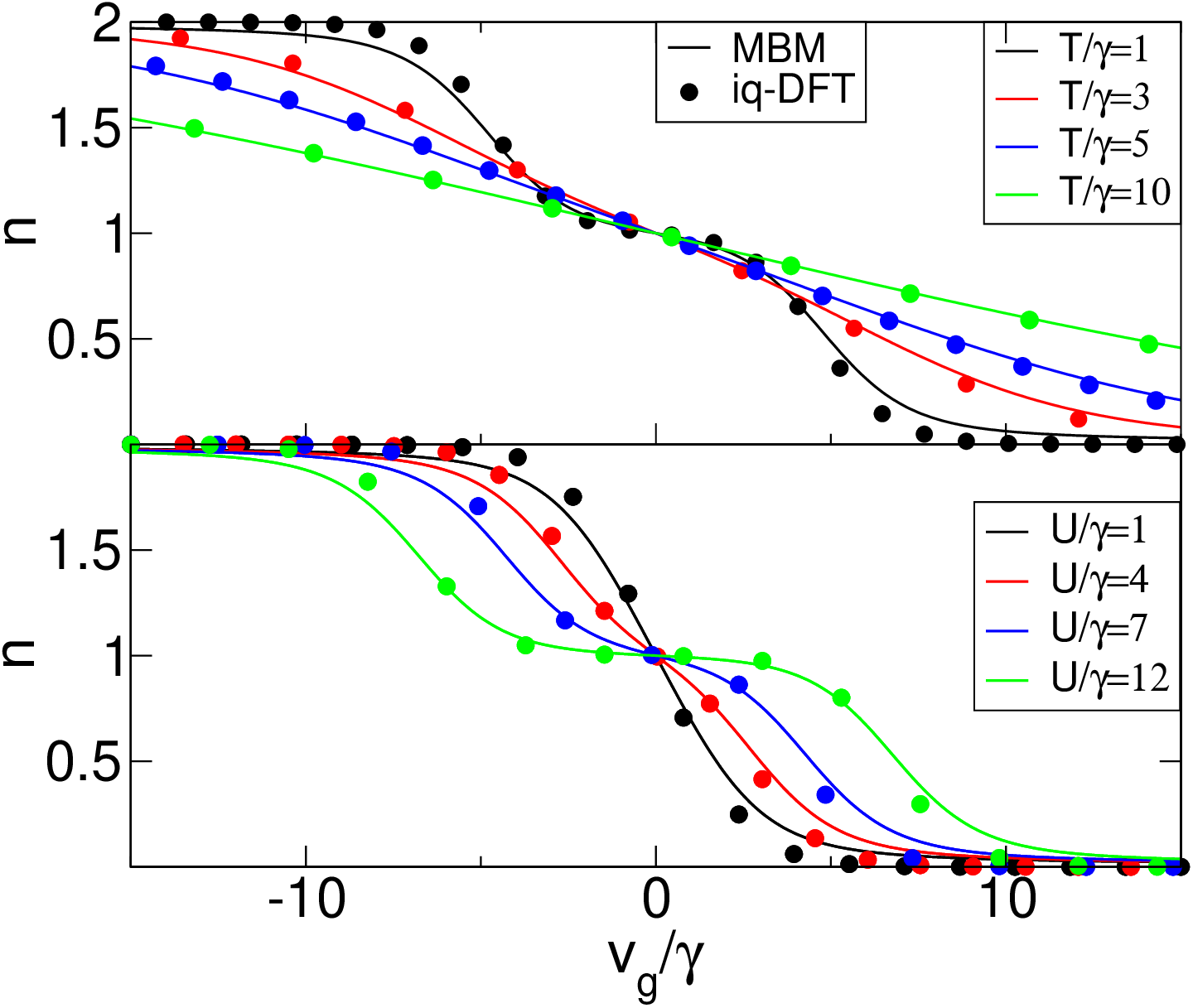}
  \caption{Equilibrium density of the SIAM as a function of the gate voltage ($v_g=v+\frac{U}{2}$)
    from the MBM and iq-DFT. In iq-DFT, the Hxc potential of the single site model has been used (see
    main text). Upper panel: density for different temperatures at fixed $U/\gamma=8$, lower panel:
    density for different interactions at fixed $T/\gamma=1$.}
  \label{fig:densities}
\end{figure}

\subsection{Numerical results}
In order to assess the accuracy of our reverse engineered approximations
for the derivatives of the iq-DFT xc potentials in comparison to the reference
MBM, we solve the DFT problem in the standard way. In the present work we use
as approximation for the Hxc (gate) potential the exact Hxc potential of the
single site model \cite{stefanucci2011towards,sobrino2019steady} given as
\be 
v_{\rm Hxc}^{\rm SSM}(n) = v_s(n)-v(n)
\label{hxc_ssm}
\ee
with $v_s(n)$ and $v(n)$ of Eqs.~(\ref{eq:vs}) and (\ref{eq:v}), respectively.
In Fig.~\ref{fig:densities}, the iq-DFT densities
as function of the gate voltage ($v_g=v+\frac{U}{2}$) are compared with the ones obtained from MBM.
As expected, this approximate Hxc potential works better as $T/\gamma$ is increased (for fixed
$U/\gamma$) while for relatively small $T/\gamma=1$ the qualitative behaviour of the density is
captured well for different interactions while quantitative differences persist.

\begin{figure}
  \includegraphics[width=\columnwidth]{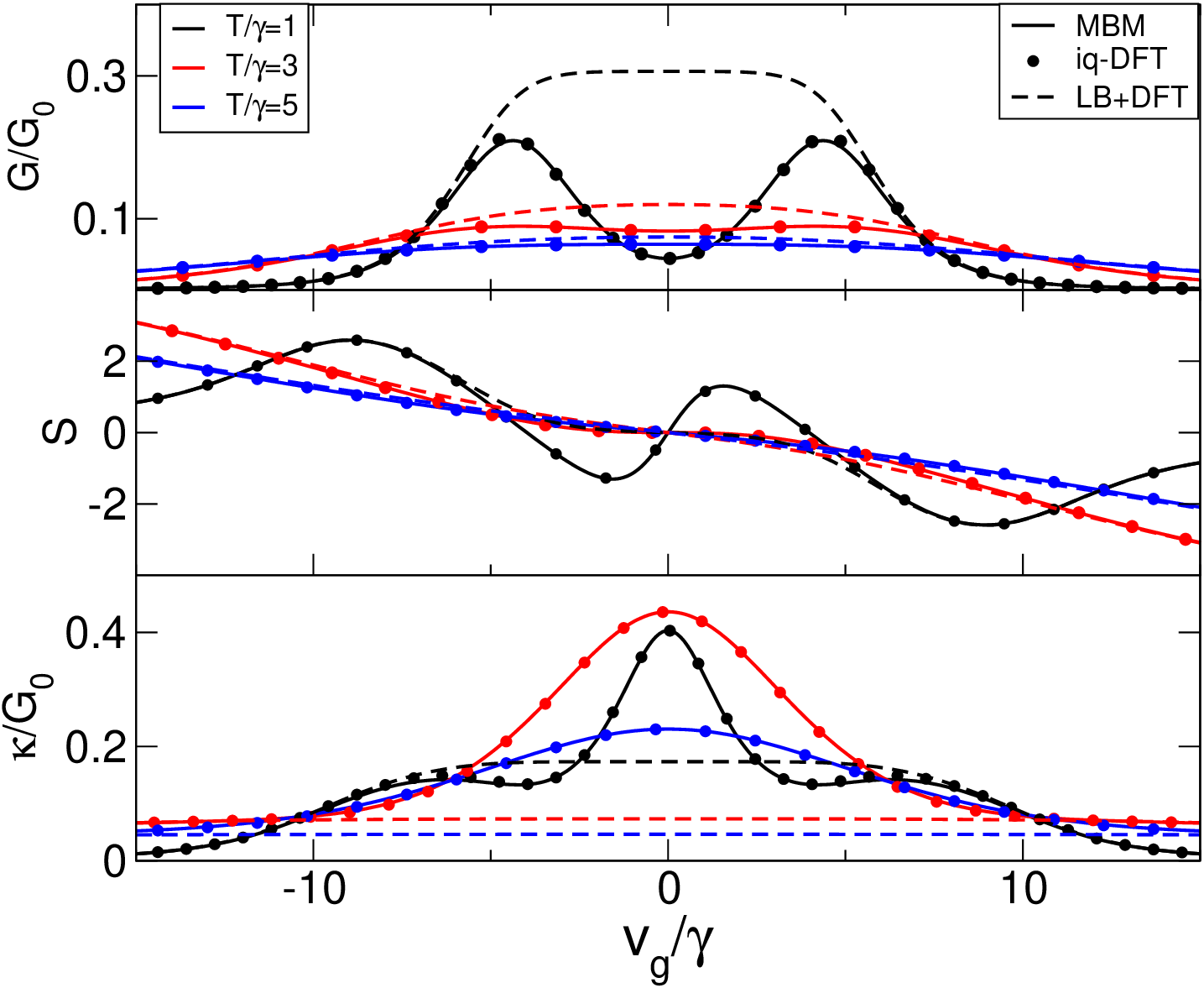}
  \caption{Transport coefficients and electronic contribution to the figure of merit  as a function of the gate voltage ($v_g=v+\frac{U}{2}$) for $U/\gamma = 8$. The iq-DFT results using the analytic reverse engineered xc derivatives Eqs.~(\ref{eq:XC_derivs_G_ijs})  are compared with those obtained directly from Eqs.~(\ref{eq:int_eqs}) when using the model spectral function of Eq.~(\ref{eq:model_spectral_function_SIAM}).}
  \label{fig:th_coeffs_varying_gate}
\end{figure}

In Fig.~\ref{fig:th_coeffs_varying_gate}, we show the linear transport coefficients for a fixed
interaction strength $U/\gamma=8$ and various temperatures as function of the gate voltage $v_g$ for the MBM, iq-DFT, and the LB-DFT approach (corresponding to the KS transport coefficients). The iq-DFT
results agree extremely well with the MBM ones highlighting the good approximation of the gate-density
relations (Eqs.~(\ref{eq:v_s_and_v_of_n})) in the range $T/\gamma>1$. On the other hand, the LB-DFT
results are only accurate in the empty orbital regime where correlations play essentially no role. 
Notice that, for $T/\gamma=1$, in the LB-DFT the Seebeck coefficient flattens around $v_g/\gamma=0$, while both iq-DFT and MBM predict a significant deviation.\cite{yang2016density,sobrino2019steady}
Note that the electrical and heat conductances are shown in units of the quantum of conductance
$G_{0}$, while the Seebeck coefficient is given in atomic units.


\begin{figure}
  \includegraphics[width=\columnwidth]{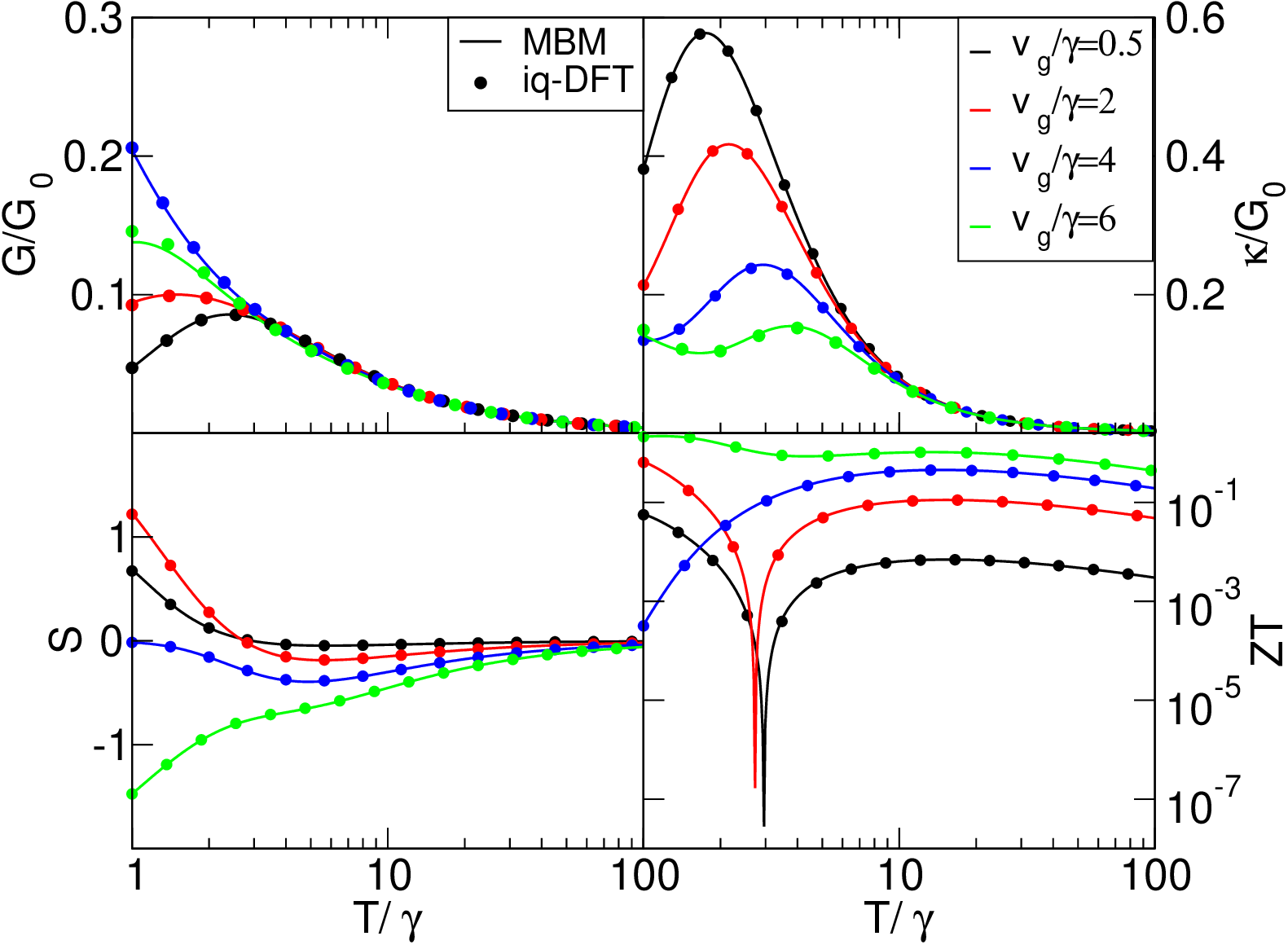}
  \caption{Transport coefficients and electronic contribution to the figure of merit  as a function of the temperature for $U/\gamma = 8$. The iq-DFT results using the analytic reverse engineered xc derivatives Eqs.~(\ref{eq:XC_derivs_G_ijs})  are compared with those obtained directly from Eqs.~(\ref{eq:int_eqs}) when using the model spectral function of Eq.~(\ref{eq:model_spectral_function_SIAM}).}
  \label{fig:th_coeffs_varying_temp}
\end{figure}

\begin{figure}
  \includegraphics[width=\columnwidth]{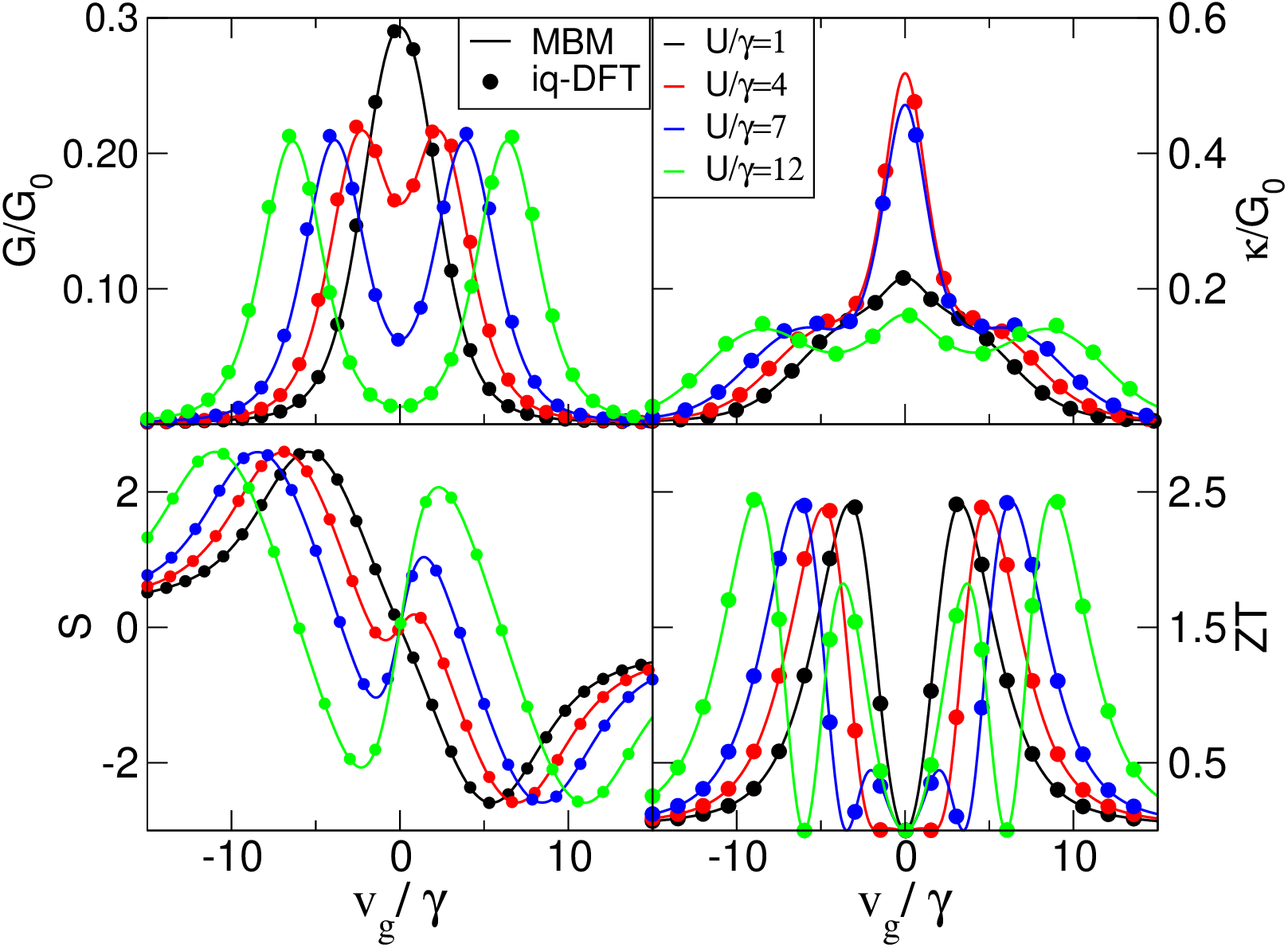}
  \caption{Transport coefficients and electronic contribution to the figure of merit as function
    of the gate voltage ($v_g=v+\frac{U}{2}$) for $T/\gamma = 1$. The iq-DFT results using the
    analytic reverse engineered xc derivatives are compared with those obtained directly from
    Eqs.~(\ref{eq:int_eqs}) when using the model spectral function of
    Eq.~(\ref{eq:model_spectral_function_SIAM}).}
  \label{fig:th_coeffs_varying_U}
\end{figure}

In Fig.~\ref{fig:th_coeffs_varying_temp}, the iq-DFT transport coefficients as well as the figure of
merit of the system $ZT=TGS^{2}/\kappa$ are
compared with those obtained from MBM for fixed gate potential as function of temperature for strong correlations $U/\gamma = 8$. As in Fig.~\ref{fig:th_coeffs_varying_gate} we observe excellent
agreement as $T/\gamma$ increases. Finally in Fig.~\ref{fig:th_coeffs_varying_U}, we show the iq-DFT transport coefficients for different interaction strength $U$ using xc parametrizations (Eq.~(\ref{eq:XC_derivs_G_ijs})).
Again we observe that for the given, fixed temperature $T/\gamma=1$, the whole range from weak
($U/\gamma=1$) to strong correlations ($U/\gamma>7$) is correctly captured in iq-DFT as compared to
the MBM. 

In terms of summary of the numerical results, we have shown that our parametrization for the
derivatives of the iq-DFT xc potentials leads to rather accurate reproduction of the linear response
transport coefficients of the MBM. There were two approximations involved in our iq-DFT approach:
(i) we used the approximate Hxc potential of Eq.~(\ref{hxc_ssm}) for the self-consistent calculation
of the density and (ii) the approximate density-potential relations (\ref{eq:v_s_and_v_of_n}) were used
to construct the xc derivatives as functionals of the equilibrium density. Both approximations (i)
and (ii) originate from the SSM and therefore it is not surprising that the corresponding iq-DFT
calculations show improved agreement with the MBM as temperature increases. We would also like to
emphasize again that the MBM approximation for the spectral function
Eq.~(\ref{eq:model_spectral_function_SIAM}) is by construction derived for the
Coulomb blockade regime
($T/T_{K}\gg1$ where $T_{K}$ is the Kondo temperature of the system). Therefore our approximation
cannot and should not be expected to accurately describe the linear transport coefficients of the
interacting system for temperatures in the Kondo regime ($T\ll T_K$). Nevertheless, our approximation
may very well serve as a first step towards the construction of improved approximations which are valid
in this regime as well. While such a construction is beyond the scope of the present study, we have
already observed that the low temperature behaviour of the Seebeck coefficient and the thermal
conductance are qualitatively correctly captured with the analytical approach. Therefore the main
corrections appear to be necessary for the electrical conductance, where ideas of the corresponding
i-DFT construction \cite{kurth2016nonequilibrium} are expected to be transferable to iq-DFT as well.

\section{Conclusions}

In this work we proposed a new density functional framework, which we call iq-DFT, to describe both
electronic and heat (energy) transport in the steady state for two-terminal (molecular) junctions
driven out of equilibrium by an external bias and/or temperature gradient between the leads, generalizing
our earlier i-DFT theory \cite{stefanucci2015steady} for steady-state transport. The foundation of iq-DFT
rests on the one-to-one correspondence between the set of three ``densities'' $(n,I,Q)$ and the set of
three ``potentials'' $(v,V,\Psi)$ which we proved for a finite bias and thermal gradient window around
equilibrium. Naturally, the corresponding KS system requires three xc potentials which need to be
approximated in practice. Unlike i-DFT, the new iq-DFT allows to calculate not only the density and steady
current but also the heat current of interacting junctions. The widely used LB-DFT formalism may be viewed
as a (crude) approximation to iq-DFT where the xc contributions to the bias as well to the $\Psi$-field
are neglected completely and the xc contribution to the local (gate) potential is independent of the
currents $I$ and $Q$. 

We developed the iq-DFT linear response formalism which allows to access all linear thermal transport
coefficients, i.e., the electrical conductance, the Seebeck coefficient, as well as the (electronic
contribution to) the thermal conductance. All these coefficients can fully and exactly be expressed in
terms of quantities accessible with iq-DFT, leading to xc corrections for all three transport
coefficients. This goes beyond i-DFT where only the electrical conductance \cite{stefanucci2015steady} 
and the Seebeck cefficient \cite{sobrino2019steady} can exactly be written in terms of quantities of
the theory. 

As a first example, we applied iq-DFT in the linear response regime to the Anderson model. From
reverse engineering of a many-body model spectral function valid in the Coulomb blockade regime,
we constructed fully analytical parametrizations of the derivatives of the iq-DFT xc potentials which
accurately reproduce the transport coefficients of the many-body model. These parametrizations are
expected to serve as a first step towards construction of approximate xc functionals beyond the Coulomb
blockade regime, in analogy to corresponding i-DFT work for the conductance \cite{kurth2016nonequilibrium}.

As any DFT framework, due to the non-interacting nature of the KS system iq-DFT can be expected to be a
numerically highly efficient scheme for the ab-initio calculation of current and heat transport through
nanoscale systems as accurate approximations for the xc functionals become available.


\section{ACKNOWLEDGMENTS}
We gratefully acknowledge useful discussions with David Jacob. We acknowledge funding  by  the  grant
“Grupos  Consolidados  UPV/EHU  del Gobierno  Vasco”  (IT1249-19).
R.D'A. acknowledges support from the Red Consolider of Spanish Government MINECO ``TowTherm'' (Grant No. MINECOG17/A01).
G.S. acknowledges financial support from MIUR PRIN 
(Grant No. 20173B72NB), from INFN through the TIME2QUEST project, and from 
Tor Vergata University through the Beyond Borders Project ULEXIEX.

\appendix*
\label{Appendix_1}
\section{Analytic expressions for the transport integrals in the SIAM}

In this Appendix we analytically evaluate the most important integrals needed both in the MBM and in
the construction of our parametrization for the derivatives of the iq-DFT xc potentials. In our MBM
for the SIAM, the many-body spectral function consists of two Lorentzians with broadening $\gamma$
($\gamma>0$) centered at $v$ and $v+U$, see
Eqs.~(\ref{eq:model_spectral_function_SIAM}). Therefore, all integrals needed to compute the
MBM density and currents have the form $\int d\omega \omega^{n}f(\omega)A(\omega)$ with $n=0,1$. 
The first integral we are interested in is
\be
\mathcal{I}_1=\int\limits_{-\infty}^\infty f(x-V/2)\frac{\gamma}{(x-x_{0})^{2}+\frac{\gamma^{2}}{4}}\mathrm{d}x,
\ee
where the Fermi function $f(z)$ can be expanded as
\cite{fetter2012quantum}
\be
f(z)=\frac{1}{1+e^{\frac{z}{T}}}  = \frac{1}{2}-\frac{\mathrm{i}}{2\pi}\sum \limits_{n=0}^\infty
\frac{1}{n+\frac{1}{2} +\mathrm{i}\frac{z}{2\pi T}}.
\label{eq:fermi_series}
\ee
Using the substitution $x = V/2 + T z$ and the abbreviations $a = \frac{x_0 - V/2}{T}$ and
$b = \frac{\gamma}{2T}$, we can write the integral as
\be
\mathcal{I}_{1} =2b \int \limits_{-\infty}^\infty \frac{\mathrm{d} z}{(\mathrm{e}^{z} + 1)[(z- a)^2 + b^2]}
\equiv 2b \int \limits_{-\infty}^\infty g(z) \mathrm{d} z . 
\ee
The integrand $g(z)$ has only single poles with non-vanishing imaginary part in the complex plane. We
therefore use the calculus of residues to compute this integral. Since $g(z)$ vanishes sufficiently
fast as $|z|\to \infty$ we can close the integration contour by a semi-circle with infinite radius
in the upper half plane (avoiding the poles on the imaginary axis). The integral then can be evaluated
as  
\be
\mathcal{I}_{1} = 4b\pi\mathrm{i} \left[ \frac{1}{\mathrm{i} 2b (\mathrm{e}^{ a+ \mathrm{i} b} + 1)}
- \sum \limits_{n=0}^\infty \frac{1}{[(2n+1) \pi \mathrm{i} - a]^2 + b^2} \right].
\ee
For the terms in the sum, we perform a fractional decomposition and then use the series representation
of the digamma function $\psi$
\be
\psi(z)=\sum \limits_{n=0}^\infty\left(\frac{1}{n+1}-\frac{1}{n+z}\right)-\gamma^{EM},
\label{eq:digamma_series}
\ee
with the Euler-Mascheroni constant $\gamma^{EM}\sim0.5772$ to obtain
\begin{eqnarray}
  \mathcal{I}_{1} &=& 2 \pi \left[\frac{1}{\mathrm{e}^{ a + \mathrm{i} b} + 1}
    - \frac{1}{2 \pi \mathrm{i}} \psi \left(\frac{1}{2} + \frac{b + \mathrm{i} a}{2 \pi}\right)
    \right. \nonumber\\
    && \left. + \frac{1}{2 \pi \mathrm{i}}  \psi \left(\frac{1}{2} + \frac{-b + \mathrm{i} a}{2 \pi} \right)\right].
\end{eqnarray}
We then apply the reflection formula
\be
\psi(1-z)=\psi(z)+\pi\cot{\left(\pi z\right)}
\ee
to the last term and, returning to the original parameters, finally arrive at
\be
\mathcal{I}_{1}(\gamma,x_{0}, V/2,T) = \pi - 2 \operatorname{Im} \left[\psi \left(\frac{1}{2} + \frac{\frac{\gamma}{2} + \mathrm{i} (x_0 - V/2)}{2 \pi T}\right) \right] .
\label{eq:I1_digamma}
\ee
In the special case $x_0 = V/2$, $\mathcal{I}_1=\pi$. 

The second integral we are interested in is 
\begin{eqnarray}
  \mathcal{I}_2&=&\int\limits_{-\infty}^\infty {\rm d}x \;
  \left(f(\frac{x-V/2}{1+\Psi/2})-f(\frac{x+V/2}{1-\Psi/2})\right) \nonumber\\
  &&\times \frac{x\gamma}{(x-x_{0})^{2}+\frac{\gamma^{2}}{4}} \;,
\end{eqnarray}
where $\Psi=\frac{T_L-T_R}{T}$. We can rewrite $\mathcal{I}_{2}$ by decomposing the second factor as
\begin{eqnarray}
  \frac{x\gamma}{(x-x_{0})^{2}+\frac{\gamma^{2}}{4}}  =
  &\frac{\gamma}{2}\left[G^{A}(x)+G^{R}(x)\right]\nonumber\\
  &- \mathrm{i}x_{0}\left[G^{A}(x)-G^{R}(x)\right]
\label{decompose}
\end{eqnarray}
with the advanced and retarded Green function
$G^{A/R}(x)=\frac{1}{x-\left(x_{0}\pm \mathrm{i}\frac{\gamma}{2}\right)}$. Noting that the
last term on the r.h.s. of Eq.~(\ref{decompose}) reduces to a Lorentzian we obtain
\be
\mathcal{I}_2=\mathcal{I}_2^{A}+\mathcal{I}_2^{R}+x_{0}\mathcal{I}_{3} \;.
\label{eq:I2_sum}
\ee
where $\mathcal{I}_{3}=\mathcal{I}_{1}\left(\gamma,x_{0}, V/2,T_L\right)-
\mathcal{I}_{1}(\gamma,x_{0}, -V/2,T_R)$ and we have defined
\bea
  \mathcal{I}_2^{A/R}&=&\frac{\gamma}{2}\lim_{r \to \infty} \int\limits_{-r}^\infty {\rm d} x \;G^{A/R}(x)
\nonumber\\
&&\times\left[f\left(\frac{x-V/2}{1+\Psi/2}\right)-f\left(\frac{x+V/2}{1-\Psi/2}\right)\right] ,
\label{eq:I2_AR_cutoff}
\eea
The integrals $\mathcal{I}_2^{A/R}$ are convergent because the
difference of the Fermi functions decays asymptotically at least as $x^{-1}$ as $|x|\to \infty$
and the Green function contributes another asymptotic $x^{-1}$ behaviour in the same limit.
Note that a lower cutoff has been explicitly introduced
in Eq.~(\ref{eq:I2_AR_cutoff}) to correctly account for the non-equivalent asymptotics of the
two Fermi functions due to their generally different temperatures (in general, $\Psi \neq 0$). .

By simple variable substitution, the integrals $\mathcal{I}_2^{A/R}$ can be written as
\begin{eqnarray}
  \lefteqn{
  \mathcal{I}_2^{A/R} = \frac{\gamma}{2}\lim_{r \to \infty} \int\limits_{\frac{-r+V/2}{1-\psi/2}}^\infty
          {\rm d} z f_T(z)} \nonumber \\
  &&\times \left( \frac{1}{z + \frac{V/2-(x_0 \pm i \gamma/2)}{1+\Psi/2}} -
  \frac{1}{z + \frac{-V/2-(x_0 \pm i \gamma/2)}{1-\Psi/2}}\right)\;\nonumber\\
  && + \frac{\gamma}{2}\lim_{r \to \infty} \int\limits_{\frac{-r-V/2}{1+\psi/2}}^{\frac{-r+V/2}{1-\psi/2}}
          {\rm d} z \; \frac{f_T(z)}{z + \frac{V/2-(x_0 \pm i \gamma/2)}{1+\Psi/2}} 
\label{eq:I2_AR_2}
\end{eqnarray}
The first contribution can now again be evaluated by closing the contour with a semicircle in
the upper half plane and summing the residues of all poles inside the contour which again leads
to digamma functions. On the other hand, the second integral becomes trivial by replacing
$f(z)$ with unity which is justified in the limit $r\to\infty$. This leads to  
\begin{eqnarray}
\mathcal{I}^{A}&=&\frac{\gamma}{2}\left[\psi\left(\frac{1}{2}+\frac{\gamma/2-\mathrm{i}(x_{0}-V/2)}{2\pi T_L}\right)\right.\nonumber\\
  && \left.-\psi\left(\frac{1}{2}+\frac{\gamma/2-\mathrm{i}(x_{0}+V/2)}{2\pi T_R}\right)\right]
\nonumber\\
&&+\frac{\gamma}{2} \log \left(\frac{1 + \Psi/2}{1-\Psi/2}\right).
\label{eq:IA}
\end{eqnarray}
and
\begin{eqnarray}
\mathcal{I}^{R}&=&\frac{\gamma}{2}\left[\psi\left(\frac{1}{2}+\frac{\gamma/2+\mathrm{i}(x_{0}-V/2)}{2\pi T_L}\right)\right.\nonumber\\
  &&\left.-\psi\left(\frac{1}{2}+\frac{\gamma/2+\mathrm{i}(x_{0}+V/2)}{2\pi T_R}\right)\right]
\nonumber\\
&&+\frac{\gamma}{2} \log \left(\frac{1 + \Psi/2}{1-\Psi/2}\right).
\label{eq:IR}
\end{eqnarray}
Using
$\operatorname{Re}\left(\psi(a+\mathrm{i}b)\right)=\frac{1}{2}\left(\psi(a+\mathrm{i}b)+\psi(a-\mathrm{i}b)\right)$ we arrive at the final result for our second integral
\begin{eqnarray}
\mathcal{I}_2&=&\gamma\operatorname{Re} \left[\psi\left(\frac{1}{2}+\frac{\gamma/2+\mathrm{i}(x_{0}-V/2)}{2\pi T_L})\right)\right]\nonumber\\
&&-\gamma\operatorname{Re} \left[ \psi\left(\frac{1}{2}+\frac{\gamma/2+\mathrm{i}(x_{0}+V/2)}{2\pi T_{R}})\right)\right]\nonumber\\
&&+x_{0}\mathcal{I}_3+ \gamma \log \left(\frac{1 + \Psi/2}{1-\Psi/2}\right).
\label{eq:I2_digamma}
\end{eqnarray}

The results for the integrals of Eqs.~(\ref{eq:I1_digamma}) and (\ref{eq:I2_digamma}) are sufficient
to analytically evaluate the density and currents for the SIAM both in the many-body model as well as
in the KS system. With these integrals we can also derive the analytical expressions for the
integrals entering the transport coefficients in the linear esponse regime. These coefficients are
\begin{subequations}
\begin{align}
  M_{11}(v) &=\frac{\gamma}{4\pi}\left.\frac{{\rm d}\mathcal{I}_3}{{\rm d}V} \right|_{\substack{V=0\\\ \Psi=0}}
  = \frac{-\gamma^{2}}{4\pi}\int f'(\omega)\frac{{\rm d}\omega}{(\omega-v)^{2}+\frac{\gamma^{2}}{4}}
\nonumber\\
&=\frac{\gamma}{4\pi^{2}T}\operatorname{Im}{\left(\mathrm{i}\psi^{(1)}( z_1)\right)}
\label{eq:M11} \\
M_{12}(v) &=\frac{\gamma}{4\pi}\left.\frac{{\rm d}\mathcal{I}_2}{{\rm d}V} \right|_{\substack{V=0\\\ \Psi=0}}
=\frac{-\gamma^{2}}{4\pi}\int f'(\omega)\frac{\omega\;{\rm d}\omega }{(\omega-v)^{2}+\frac{\gamma^{2}}{4}}
\nonumber\\
&=\frac{\gamma}{4\pi^{2}T}\operatorname{Im}{\left(z_{0}\psi^{(1)}( z_1)\right)}
\label{eq:M12}
\\
M_{22}(v) &=\frac{\gamma}{4\pi}\left.\frac{d\mathcal{I}_2}{d\Psi} \right|_{\substack{V=0\\\ \Psi=0}}
= \frac{-\gamma^{2}}{4\pi}\int  f'(\omega)\frac{\omega^{2}\;{\rm d}\omega}{(\omega-v)^{2}+\frac{\gamma^{2}}{4}}\nonumber\\
&=-\frac{\gamma^{2}}{8\pi^{2}T}\operatorname{Re}{\left(z_{0}\psi^{(1)}(z_1)\right)}+vM_{12}+\frac{\gamma^{2}}{4\pi},
\label{eq:M22}
\end{align}
\label{eq:Mij}
\end{subequations}
where $z_0=\frac{\gamma}{2}+\mathrm{i}v$, $z_{1}=\frac{1}{2}+\frac{z_{0}}{2\pi T}$, and
$\psi^{(1)}(z)$ is the trigamma function.\cite{AbramowitzStegun:65}


%

\end{document}